\documentclass[aps,prl,reprint,superscriptaddress]{revtex4-1}
\usepackage{graphicx}
\usepackage{color,amsmath}
\usepackage[normalem]{ulem}
\usepackage{hyperref}
\usepackage{lipsum}

\newcommand{\vpg}{V_{\rm{pg}}}
\newcommand{\vc}{V_{\rm{c}}}
\newcommand{\vsd}{V_{\rm{sd}}}
\newcommand{\bpl}{B_{\parallel}}
\newcommand{\bperp}{B_{\perp}}
\newcommand{\btr}{B_{\rm{t}}}
\newcommand{\br}{B_{\rm{r}}}
\newcommand{\Ec}{E_{\rm{c}}}
\newcommand{\Eo}{E_{\rm{0}}}

\newcommand{\Se}{S_{\rm{e}}}

\newcommand{\So}{S_{\rm{o}}}

\newcommand{\Savg}{\langle S \rangle}

\newcommand{\Aavg}{\langle \tilde{A} \rangle}

\begin{document}

\title{Coherent transport through a Majorana island \\ in an Aharonov-Bohm interferometer}

\author{A.~M.~Whiticar}
\altaffiliation{These authors contributed equally to this work.}
\affiliation{Center for Quantum Devices, Niels Bohr Institute, University of Copenhagen and Microsoft Quantum Lab Copenhagen, Universitetsparken 5, 2100 Copenhagen, Denmark}

\author{A.~Fornieri}
\altaffiliation{These authors contributed equally to this work.}
\affiliation{Center for Quantum Devices, Niels Bohr Institute, University of Copenhagen and Microsoft Quantum Lab Copenhagen, Universitetsparken 5, 2100 Copenhagen, Denmark}

\author{E.~C.~T.~O'Farrell}
\affiliation{Center for Quantum Devices, Niels Bohr Institute, University of Copenhagen and Microsoft Quantum Lab Copenhagen, Universitetsparken 5, 2100 Copenhagen, Denmark}

\author{A.~C.~C.~Drachmann}
\affiliation{Center for Quantum Devices, Niels Bohr Institute, University of Copenhagen and Microsoft Quantum Lab Copenhagen, Universitetsparken 5, 2100 Copenhagen, Denmark}

\author{T.~Wang}
\affiliation{Department of Physics and Astronomy and Microsoft Quantum Lab Purdue, Purdue University, West Lafayette, Indiana 47907 USA}
\affiliation{Birck Nanotechnology Center, Purdue University, West Lafayette, Indiana 47907 USA}

\author{C.~Thomas}
\affiliation{Department of Physics and Astronomy and Microsoft Quantum Lab Purdue, Purdue University, West Lafayette, Indiana 47907 USA}
\affiliation{Birck Nanotechnology Center, Purdue University, West Lafayette, Indiana 47907 USA}

\author{S.~Gronin}
\affiliation{Department of Physics and Astronomy and Microsoft Quantum Lab Purdue, Purdue University, West Lafayette, Indiana 47907 USA}
\affiliation{Birck Nanotechnology Center, Purdue University, West Lafayette, Indiana 47907 USA}

\author{R.~Kallaher}
\affiliation{Department of Physics and Astronomy and Microsoft Quantum Lab Purdue, Purdue University, West Lafayette, Indiana 47907 USA}
\affiliation{Birck Nanotechnology Center, Purdue University, West Lafayette, Indiana 47907 USA}

\author{G.~C.~Gardner}
\affiliation{Department of Physics and Astronomy and Microsoft Quantum Lab Purdue, Purdue University, West Lafayette, Indiana 47907 USA}
\affiliation{Birck Nanotechnology Center, Purdue University, West Lafayette, Indiana 47907 USA}

\author{M.~J.~Manfra}
\affiliation{Department of Physics and Astronomy and Microsoft Quantum Lab Purdue, Purdue University, West Lafayette, Indiana 47907 USA}
\affiliation{Birck Nanotechnology Center, Purdue University, West Lafayette, Indiana 47907 USA}
\affiliation{School of Materials Engineering, Purdue University, West Lafayette, Indiana 47907 USA}
\affiliation{School of Electrical and Computer Engineering, Purdue University, West Lafayette, Indiana 47907 USA}

\author{C.~M.~Marcus}
\email[email: ]{marcus@nbi.ku.dk}
\affiliation{Center for Quantum Devices, Niels Bohr Institute, University of Copenhagen and Microsoft Quantum Lab Copenhagen, Universitetsparken 5, 2100 Copenhagen, Denmark}

\author{F.~Nichele}
\email[email: ]{fni@ibm.zurich.com}
\altaffiliation{Present address: IBM Research - Zurich, Sumerstrasse 4, 8803 Rschlikon, Switzerland.}
\affiliation{Center for Quantum Devices, Niels Bohr Institute, University of Copenhagen and Microsoft Quantum Lab Copenhagen, Universitetsparken 5, 2100 Copenhagen, Denmark}
\date{\today}

\maketitle

\textbf{\boldmath 
Majorana zero modes are leading candidates for topological quantum computation due to non-local qubit encoding and non-abelian exchange statistics. Spatially separated Majorana modes are expected to allow phase-coherent single-electron transport through a topological superconducting islands via a mechanism referred to as teleportation. Here we experimentally investigate such a system by patterning an elongated epitaxial InAs-Al island embedded in an Aharonov-Bohm interferometer. With increasing parallel magnetic field, a discrete sub-gap state in the island is lowered to zero energy yielding persistent 1$e$-periodic Coulomb blockade conductance peaks ($e$ is the elementary charge). In this condition, conductance through the interferometer is observed to oscillate in a perpendicular magnetic field with a flux period of $h/e$ ($h$ is Planck's constant), indicating coherent transport of single electrons through the islands, a signature of electron teleportation via Majorana modes.}

\section{Introduction}

\begin{figure*}
	\includegraphics[width=2\columnwidth]{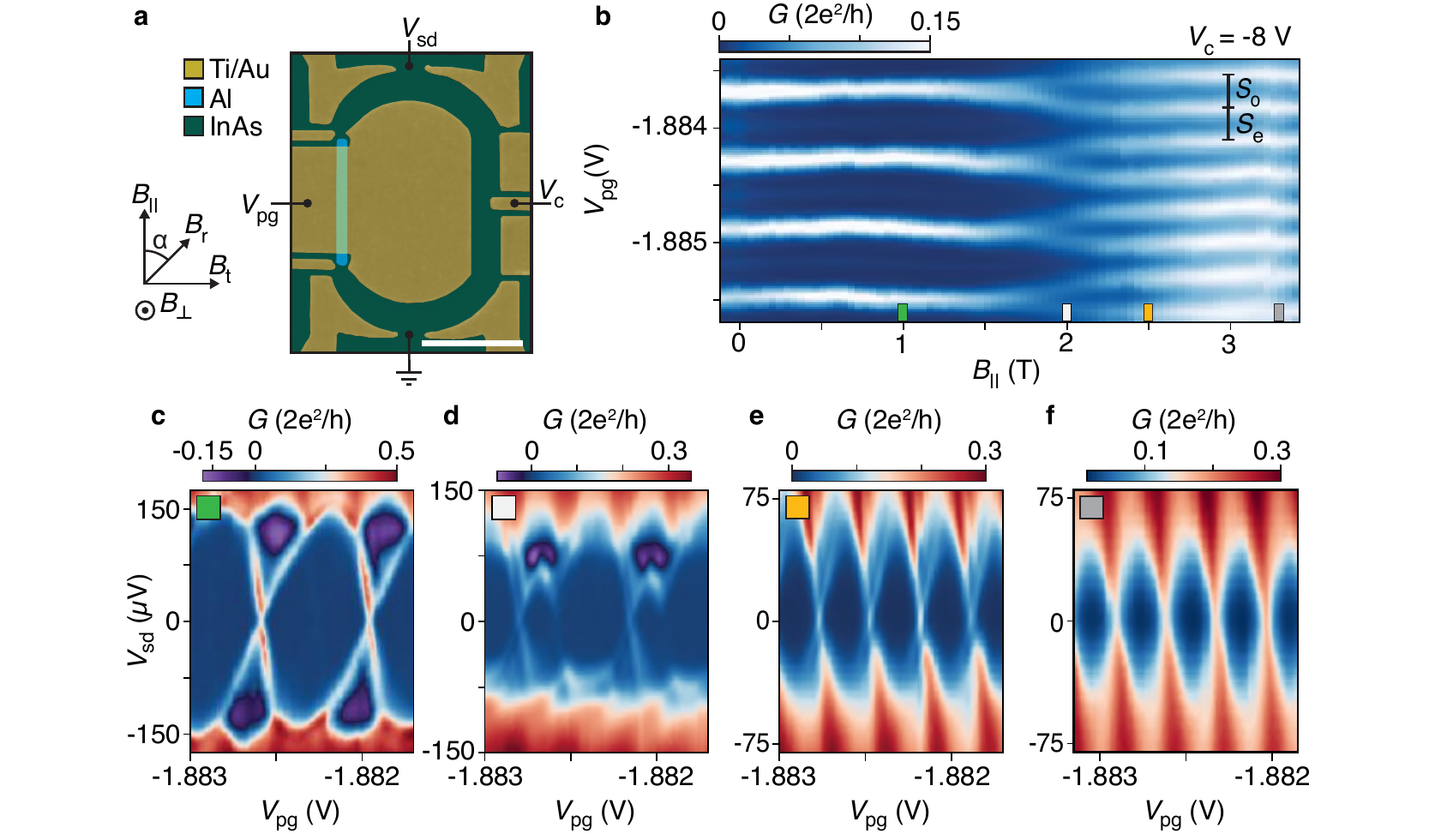}
	\caption{\textbf{Majorana island interferometer.} \textbf{a}, False-colour electron micrograph of the Majorana island interferometer where an Al wire (light blue) is embedded in a normal conducting Aharonov-Bohm interferometer (green) defined by Ti/Au gates (yellow). The gate voltage $\vpg$ defines both the Majorana island and the interferometer center, and controls the electron occupancy. The gate voltage $\vc$ controls the resistance of the reference arm and $\vsd$ is the source-drain dc bias voltage. Magnetic field directions are shown with $\alpha$ denoting the in-plane angle measured with respect to the wire direction. Scale bar, 1~$\mu\rm{m}$. \textbf{b}, Zero-bias differential conductance $G$ as a function of $\bpl$ and $\vpg$. $\Se$ ($\So$) is the even (odd) CB peak spacing. \textbf{c-f}, Differential conductance $G$ as a function of $\vsd$ and $\vpg$ showing Coulomb diamonds for $\bpl$ = 1 T (\textbf{c}), 2 T (\textbf{d}), 2.5 T (\textbf{e}), and 3.3 T (\textbf{f}). The measurements shown in panels \textbf{b-f} were taken with the reference arm closed.
	}
 	\label{fig1}
\end{figure*}

Initial experiments reporting signatures of Majorana zero modes (MZMs) in hybrid superconductor-semiconductor nanowires focussed on zero-bias conductance peaks (ZBPs) using local tunnelling spectroscopy~\cite{Mourik2012b,Deng2016,Nichele2017,Zhang2018}. Subsequently, Majorana islands provided additional evidence of MZMs based on nearly 1$e$-spaced Coulomb blockade (CB) peaks~\cite{Albrecht2016b}, and indicated a Rashba-like spin orbit coupling with the spin-orbit field lying in-plane, perpendicular to the wire axis ~\cite{OFarrell2018}.  Under some circumstances, these signatures can be mimicked by trivial modes~\cite{Chiu2017,Moore2018d,prada2019andreev}, motivating a new generation of experiments that explicitly probe non-local properties, which are more difficult to mimic. For instance, non-locality of MZMs was recently investigated by measuring the energy splitting induced by the interaction of a quantum dot and a zero-energy state in a hybrid nanowire~\cite{Deng2018}.

Non-locality can also be accessed by interferometric measurements of a Majorana island, where CB couples separated MZMs and fixes fermion parity~\cite{Fu2010,Sau2015a,Vijay2016,Hell2018b,Drukier2018}. In the topological regime, a Majorana island can coherently transfer a single-electron between its two ends through MZMs~\cite{Fu2010,Sau2015a}. To demonstrate the effect, a Majorana island can be embedded in the arm of an Aharonov-Bohm (AB) interferometer. If single-electron transport in both the reference arm and the Majorana island is coherent, conductance through the interferometer is expected to show oscillations with a flux period $h/e$~\cite{Aharonov1959,Fu2010}. In addition, interferometry offers a way to distinguish between localized trivial modes and MZMs~\cite{Sau2015a,Hell2018b}. This technique was used to investigate coherent transport in semiconductor quantum dots~\cite{Yacoby1996,Schuster1997a,Avinun-Kalish2005,Edlbauer2017}.

In this work, we study coherent single-electron transport through a Majorana island with an AB interferometer. At high magnetic fields, we observe $1e$ periodic CB peaks due to a discrete zero-energy state. In this case, we observe the conductance through the interferometer to oscillate periodically for a flux $h/e$ piercing the AB ring, while oscillations are suppressed when the island shows a $2e$ periodicity or is in the normal state. The observation of conductance oscillations in the $1e$ regime indicates coherent single-electron transport through the Majorana island, as predicted for electron teleportation mediated by MZMs~\cite{Fu2010}.

\begin{figure*}
	\includegraphics[width=2\columnwidth]{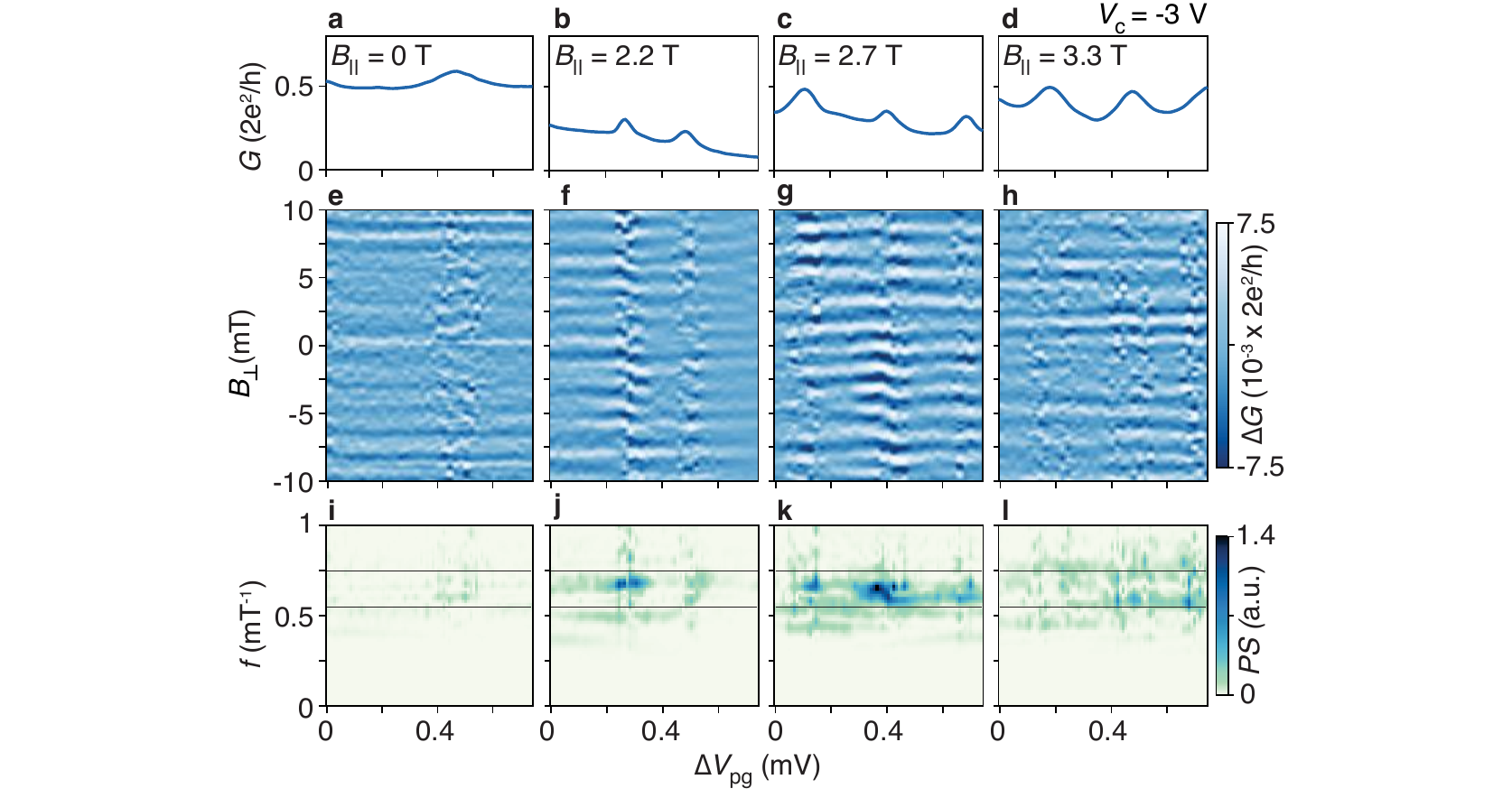}
	\caption{\textbf{\boldmath{$h/e$} periodic conductance oscillations}. Magnetoconductance for parallel field values $\bpl$ = 0, 2.2, 2.7, and 3.3~T (left to right). \textbf{a-d}, Zero-bias differential conductance $G(\bperp=0$) versus gate voltage $\vpg$ used to control electron occupation. \textbf{e-h}, Conductance $\Delta G$ as a function of $\vpg$ and perpendicular magnetic field $\bperp$ controlling the flux in the interferometer with corresponding power spectrums in \textbf{i-l}. The solid black lines indicate the frequency window bounding the Aharonov-Bohm oscillations (see methods). $\Delta G$ is the conductance with a subtracted slowly varying background. $\Delta \vpg = 0 $ corresponds to $\vpg = $ -1.896 V.
	}
	\label{fig2}
\end{figure*}

\section{Results}

\subsection{Majorana island interferometer.}

Devices were fabricated using an InAs two-dimensional electron gas (2DEG) heterostructure covered by 8 nm of epitaxially grown Al~\cite{Shabani2016a}. The bare 2DEG (without Al) showed a phase coherence length of  $l_{\phi} \sim 4~\mu \rm{m}$ (see Supplementary Figure~1). Figure \ref{fig1}a shows a micrograph of device 1 with a 1.2~$\mu$m long and  0.1~$\mu \rm{m}$ wide superconducting Al stripe formed by wet etching. Ti/Au top-gates were evaporated on top of a 25 nm HfO$_2$ dielectric grown by atomic layer deposition. We studied two lithographically similar interferometers with circumferences $L_{\rm loop}$ of 5.6~$\mu$m for device 1 and 5~$\mu$m for device 2.

Applying a negative voltage, $\vpg$, to the central gate serves two purposes. It depletes the 2DEG surrounding the Al wire to form both the Majorana island and the AB ring center and also adjusts the chemical potential and charge occupancy of the island. Energizing all exterior gates confines the 2DEG into an AB interferometer by connecting the Majorana island to a normal conducting reference arm. The resistance of the reference arm was independently tuned by a negative gate voltage $\vc$. A source-drain bias voltage ($\vsd$) was applied to one lead and the resulting current and four-terminal voltage was recorded. The in-plane magnetic fields $\bpl$ and $\btr$, and perpendicular field, $\bperp$, were controlled by a three-axis vector magnet.

At low temperatures, tunneling of single electrons onto a Majorana island with a superconducting gap $\Delta$ is suppressed by CB, except at charge degeneracies. When the lowest sub-gap state energy, $\Eo$, exceeds the charging energy $\Ec$, ground-state degeneracies only occur between  even-occupied states, resulting in 2$e$-periodic CB conductance peaks~\cite{Hekking1993a}. Odd-occupied ground states are lowered into the accessible spectrum by a Zeeman field, resulting in even-odd CB peak spacing when $ 0< \Eo <  \Ec$. The difference in peak spacings between even and odd states,  $S = \Se - \So $, is proportional to $\Eo$~\cite{Albrecht2016b} (see Fig.~\ref{fig1}b). For well-separated MZMs, $\Eo$ tends exponentially toward zero, yielding 1$e$ periodic CB peaks with a discrete zero-bias state at consecutive charge degeneracy point~\cite{VanHeck2016,Albrecht2016b}. Both observations are necessary for a MZM interpretation. When MZMs are not widely separated, CB peak spacings oscillate with field and chemical potential~\cite{OFarrell2018,Albrecht2016b,Chiu2017}.

\subsection{Coulomb blockade spectroscopy.}
We first investigated the Majorana island without interferometry by depleting a segment of the reference arm (see Fig.~\ref{fig1}a). Figure \ref{fig1}b shows zero-bias differential conductance $G = dI/dV$ of the island as a function of parallel magnetic field $\bpl$ and gate voltage $\vpg$, which controls the electron occupancy and chemical potential of the island. CB peaks are 2$e$ periodic at zero field and split around 2~T, becoming 1$e$ periodic as the sub-gap state moves toward zero energy (see Fig. \ref{fig3}a for peak spacing analysis). Performing CB spectroscopy, that is, measuring $G$ as a function of both source-drain bias $\vsd$ and $\vpg$ reveals Coulomb diamonds (Fig.~\ref{fig1}c-f). At low $\bpl$, diamonds are 2$e$ periodic with distinct negative differential conductance (Fig.~\ref{fig1}c), which transition to an even-odd peak spacing difference at moderate fields (Fig.~\ref{fig1}d), similar to previous work on superconducting Coulomb islands~\cite{Tuominen1992,Eiles1993,Hekking1993a,Higginbotham2015b,Albrecht2016b,Shen2018,OFarrell2018}.  At high fields, the 1$e$ periodic diamonds show a discrete ZBP for consecutive charge degeneracy points that is well separated from the superconducting gap (Fig.~\ref{fig1}e). This sub-gap feature remained at zero bias until the superconducting gap closure, and persists for 3 mV in $\vpg$, corresponding to an energy range of 0.8 meV. For systems with strong spin-orbit interaction, a helical gap of $g\mu_{\rm B}\bpl$ is expected, where $\mu_{\rm B}$ and $g$ are the Bohr magneton and electron g-factor, respectively. We estimate a helical gap of $\sim 0.6$~meV at $\bpl = 2.5$~T for a g-factor of 4, comparable to the span of the ZBP in $\vpg$. The stability of the zero-bias state in both magnetic field and in chemical potential is consistent with the MZM picture (see Supplementary Figure~2) \cite{Albrecht2016b}, however, the observation of coherent single-electron transport is needed to draw conclusions about non-locality. Below, we additionally show that the zero-bias state is sensitive to rotations of the in-plane field. The magnitude of $\bpl$ where 1$e$ periodicity is observed is in agreement with ZBPs measured in tunneling spectroscopy in InAs 2DEGs~\cite{Nichele2017}. In contrast, as a function of $\btr$ the peak spacing showed an oscillating even-odd periodicity, and no consecutive ZBPs were observed (see Supplementary Figure~3d-f), as expected for extended modes in the wire~\cite{OFarrell2018,prada2019andreev,stanescu2013dimensional}. The normal state 1$e$ regime of the Majorana island appears above $\bpl\sim3$ T with $\Ec =80~\mu$eV (Fig.~\ref{fig1}f), where no discrete ZBPs are observed (see Supplementary Figure~2 and 3a-c).

\subsection{Interferometry and coherent single-electron transport.}
The reference arm of the AB interferometer was connected to the Majorana island by tuning $\vc$ from -8~V to -3~V while $\vpg$ was compensated. This lifted the overall conductance by opening a path through the reference arm (see Fig.~\ref{fig2}a-d). Figure \ref{fig2}e-h shows the conductance $\Delta G$ through the full interferometer (with smooth background subtracted; see methods) as a function of $\bperp$ and gate voltage $\vpg$, which control the flux in the interferometer and occupancy of the island, respectively. Figure \ref{fig2}e shows small oscillations in $\Delta G(\bperp)$ at $\bpl = 0~\rm{T}$ for the 2$e$ periodic peaks. For $\bpl = 2.2$~T, where the peak spacing is even-odd (Fig.~\ref{fig2}f), the conductance showed moderate oscillations with a period $\Delta \bperp = 1.5$ mT. This periodicity corresponds to a single flux quantum $h/e$ threading an area of $\sim 2.7 ~\mu\rm{m}^2$, consistent with the ring center area defined by $\vpg$. This indicates coherent $1e$ transport through both the reference arm and the Majorana island. At $\bpl = 2.7$~T, the CB peak spacing is uniformly 1$e$, and oscillation amplitude is maximal (see Figs.~\ref{fig2}g). When the Majorana island is driven normal, $\bpl > 3$~T, conductance oscillations are reduced, becoming comparable to the low field oscillations (Fig.~\ref{fig2}h). The appearance of strong $h/e$ periodic conductance oscillations in the 1$e$ regime of the island is a key experimental signature of electron teleportation.

The strength and periodicity of the oscillations are examined more quantitatively using Fourier power spectrum (PS) analysis (see methods). In Figs.~\ref{fig2}i-l, the PS of $\Delta G(\bperp)$ are shown. Increasing $\bpl$ led to a peak appearing around $f = 0.65 ~\rm{mT}^{-1}$, the frequency expected for AB interference. The PS is maximized in the 1$e$ regime. To quantify the oscillations amplitude, $\Aavg$, we average the integrated PS  (see methods).

\begin{figure}
	\includegraphics[width=1\columnwidth]{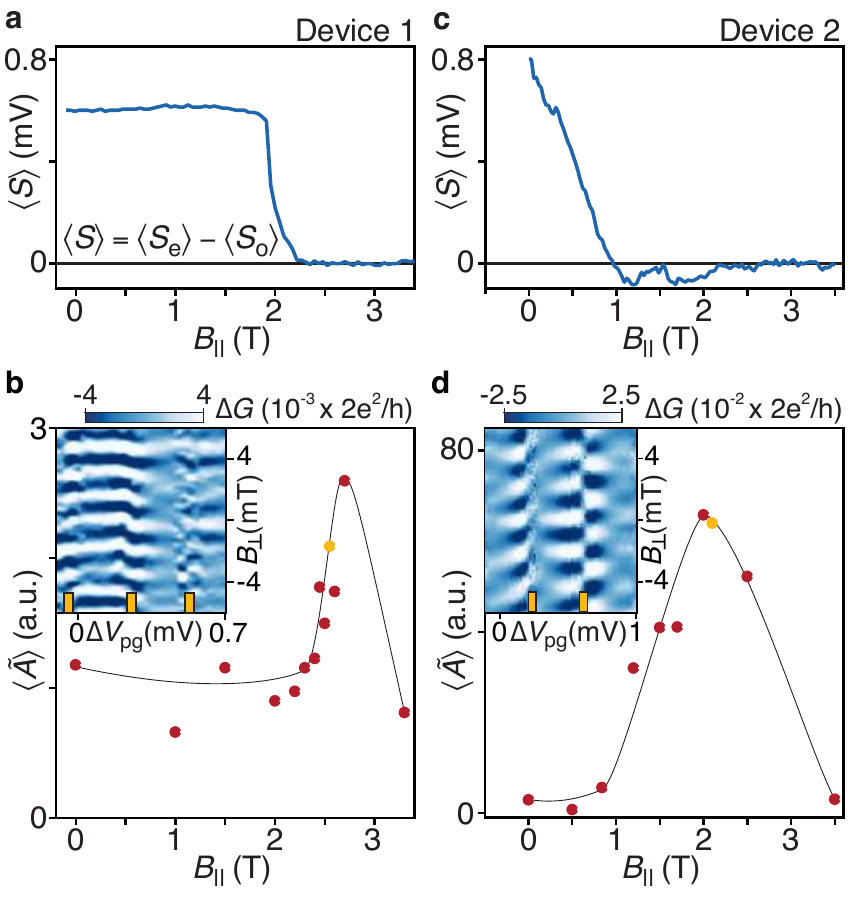}
	\caption{ \textbf{Coherent single-electron transport} \textbf{a,c}, Peak spacing difference $\Savg$ as a function of parallel magnetic field $\bpl$ for devices 1 and 2. \textbf{b,d}, Aharonov-Bohm oscillation amplitude $\Aavg$ as a function of $\bpl$. The solid line is a  guide to the eye. Insets show characteristic magnetoconductance $\Delta G$ as a function of gate voltage $\vpg$ controlling electron occupancy and perpendicular magnetic field $\bperp$ controlling the magnetic flux in the interferometer in the 1$e$ regime (indicated by yellow markers in the main panel). Yellow ticks show CB peak positions. $\Delta \vpg = 0$ corresponds to  $\vpg = $ -1.896 V and -0.945 V for \textbf{b} and \textbf{d}, respectively.
	}
	\label{fig3}
\end{figure}

We next correlate the $\bpl$ dependence of the oscillations amplitude, $\Aavg(\bpl)$, with the $\bpl$ dependence of the lowest sub-gap state, $\Eo(\bpl)$, of the island. The sub-gap energy is found from the difference between even and odd CB peak spacings, averaged separately, $\Savg = \langle \Se \rangle - \langle \So \rangle$ (see Fig. \ref{fig1}b). In Figure~\ref{fig3}a, $\Savg$ remains constant as a function of $\bpl$ (indicating 2$e$ transport) until a sub-gap state moves below $\Ec$, reaching zero at 2.2~T without overshoot (as expected for well separated MZMs in a long wire \cite{VanHeck2016,Albrecht2016b}). At low fields, where the CB periodicity is 2$e$, the oscillation amplitude $\Aavg$ is small (Fig.~\ref{fig3}b). When $\Savg$ approaches zero at high fields ($\bpl>2 $~T), $\Aavg$ exhibits a sharp increase that coincides with the 2$e$ to 1$e$ transition. Above 3~T, the device is in the normal state and $\Aavg$ returned to the low value found in the 2$e$ regime. This comparison shows that the oscillation amplitude is correlated with the energy of the lowest subgap state, and is maximal in the 1$e$ superconducting regime, as expected for electron teleportation.
 
Figure \ref{fig3}c,d shows a similar study for device 2. In Fig. \ref{fig3}c, $\Savg$ shows strong even-odd below 1 T, fluctuates around $\Savg = 0$ between 1-2~T, then settles to 1$e$ ($\Savg=0$) above 2~T. CB spectroscopy reveals a discrete state that oscillates around zero bias in both $\bpl$ and $\vpg$ without forming a stable $1e$ periodic zero-bias peak (see Supplementary Figure~5). These oscillations about zero energy are compatible with hybridized Majorana modes due to wavefunction overlap resulting from the finite island length~\cite{prada2019andreev,stanescu2013dimensional}. This overlap causes an energy splitting that oscillates both in field and chemical potential~\cite{OFarrell2018,Albrecht2016b,Shen2018}. Figure \ref{fig3}d shows that phase coherent transport first appears above 1~T and $\Aavg$ gradually increases until reaching a maximum amplitude for 1$e$ peak spacing at 2.1~T, before diminishing in the normal state. In comparison to device 1, phase coherent transport appears when $\Savg$ oscillations about zero, suggesting that hybridized Majorana modes characterized by an extended wavefunction may also contribute to coherent transport. The absence of energy oscillations may distinguish non-local MZMs from hybridized Majorana modes (see Fig. \ref{fig3}a,c). We attribute the reduced oscillation amplitude in device 1 to result from the longer loop length $L_{\rm loop}>l_{\phi}$, leading to increased dephasing in the reference arm.

We observe that conductance oscillations measured on opposite sides of a CB peak in the $1e$ regime are out of phase (see yellow ticks in the insets of Fig. \ref{fig3}) indicating a transmission phase shift of $\pi$ is acquired when the parity of the island is flipped. This demonstrates interferometric detection of island parity, which offers a way to detect MZM parity as proposed by several recent topological qubit schemes ~\cite{Karzig2017b,Plugge2017,Vijay2016}. In some cases, we found that the the phase shift is restored through the CB valley, such that the same sides of adjacent CB peaks have the same phase. What determines whether there is phase recovery in the CB valley is not currently understood (see Supplementary Figure~6)~\cite{Drukier2018,Schuster1997a,Avinun-Kalish2005,Edlbauer2017}.

\begin{figure}
	\includegraphics[width=1\columnwidth]{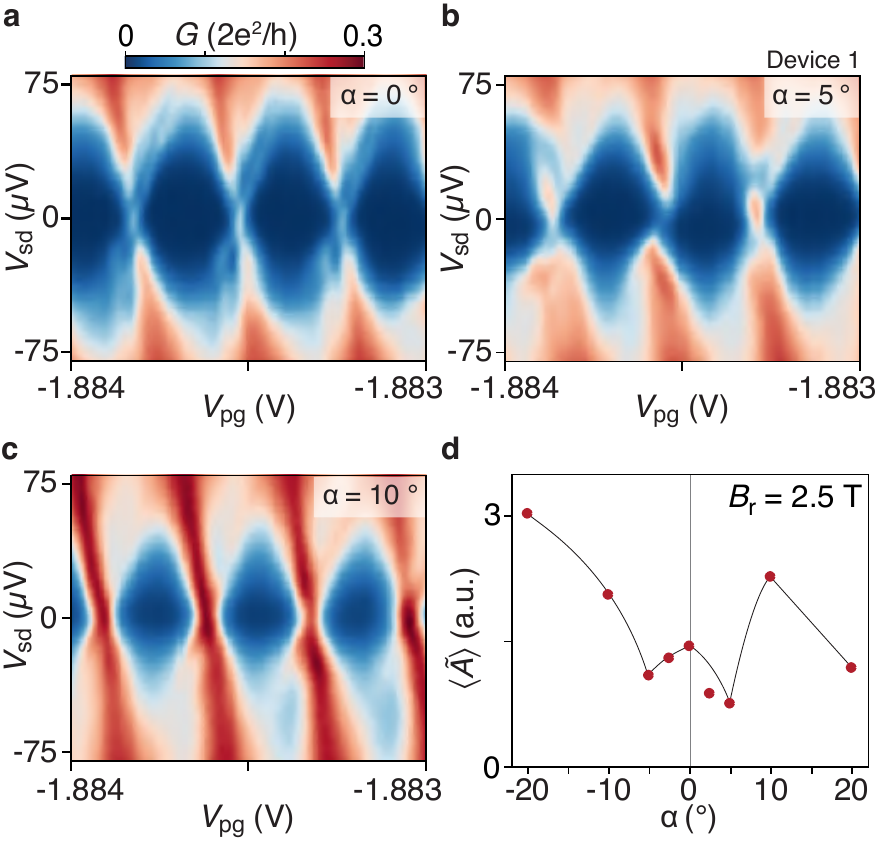}
	\caption{\textbf{In-plane magnetic field rotations.} \textbf{a-c}, Differential conductance $G$ as a function of gate voltage $\vpg$ controlling electron occupancy and source-drain bias voltage $\vsd$ showing Coulomb diamonds for in-plane rotation angles of $\alpha$  = 0 (\textbf{a}), 5$^\circ$ (\textbf{b}), and 10$^\circ$ (\textbf{c}) with $\alpha = 0$ corresponding to $\bpl = 2.5 $~T. \textbf{d}, Oscillation amplitude $\Aavg$ as a function of in-plane rotation angle $\alpha$ for $\br = 2.5$~T. The solid line is a guide to the eye.
	}
	\label{fig4}
\end{figure}

The angular dependence of the in-plane magnetic field is investigated by fixing the field magnitude $\br = 2.5~\rm{T}$ and rotating the field by an angle $\alpha$ (see Fig. \ref{fig1}). Theoretically, a rotation of the in-plane field towards the Rashba field direction is expected to close the topological gap~\cite{Osca2014}. Figure \ref{fig4}a shows 1$e$ periodic Coulomb diamonds at $\bpl = 2.5$~T with a discrete ZBP at each charge degeneracy point (similar to Fig.~\ref{fig1}e). Rotating by an angle $\alpha = 5^{\circ}$ lifted the discrete state from zero energy, leading to even-odd peak spacing; at $\alpha = 10^{\circ}$,  $1e$ periodicity is recovered, though without a discrete ZBP. The observed sensitivity of the zero-energy state to in-plane field rotation is consistent with MZMs~\cite{Osca2014}. 

Small rotations ($| \alpha |<7.5 ^{\circ}$) reduced the oscillation amplitude, $\Aavg$, as expected for even-odd periodicity (see Fig. \ref{fig3}). However, at larger angles ($| \alpha |>10 ^{\circ}$) where the discrete ZBP is absent, a strong interference signal is observed  (Fig.~\ref{fig4}d). The observation of coherent transport in the absence of a discrete zero-energy state suggests trivial extended modes are also phase coherent over the length of the island. Therefore, the additional information provided by bias spectroscopy is needed to distinguish teleportation from other coherent transport mechanisms, as shown in Figs.~\ref{fig4}a-c.

We further studied the effect of different magnetic field direction. The results are shown in Supplementary Figure~7. In summary, all three axes showed coherent transport, with oscillation amplitude first increasing as $\Savg$ approached zero. This shows that the oscillation amplitude is dictated by the energy $\Eo$ in all field directions and indicates that interference is not unique to a parallel magnetic field.

Finally, we comment on the physical mechanism that correlates the oscillation amplitude to the energy of $\Eo$. At low fields, the Majorana island favours an even parity where transport of electrons occurs as two sequential tunneling events on either end of the island~\cite{Hekking1993a,Eiles1993}. The two electrons acquire the condensate phase when forming a Cooper pair, which suppresses single-electron coherence. At moderate fields, a discrete sub-gap state is brought below $\Ec$ and a single-electron transport channel is opened, allowing coherent resonant tunneling through the Majorana island. When the discrete state is brought to zero-energy, the contribution of coherent transport is increased due to electron teleportation. Finally, in the normal state, we interpret the reduction in interference signal to reflect the short coherence length in the diffusive Al wire. 

In conclusion, we report signatures of single-electron teleportation via non-local MZMs using AB interference in combination with spectroscopy of a discrete zero-energy state. Our results also reveal that coherent transport by topologically trivial modes extending over the full length of the Majorana island are allowed. These extended trivial modes may be precursors of topological states in a finite length system that could transition into non-local MZMs by adjusting experimental parameters~\cite{stanescu2013dimensional}. We have shown that interferometry accompanying bias spectroscopy revealing stable 1$e$ periodic CB in magnetic field and chemical potential can discriminate non-local MZMs from extended modes (that display characteristic energy oscillations). Increasing the wire length to greatly exceed the diffusive coherence length $\xi = \sqrt{\xi_0~l_e}\sim 1~\mu$m (for $\Delta =  75~\mu$eV at $\bpl = 2.5$~T), where $\xi_0$ is the clean coherence length and $l_e \sim~300$ nm is the semiconducting mean free path  will suppress $1e$ transport via trivial extended modes ~\cite{tinkham2004introduction}. The observation of coherent transport through the island rules out localized ABS at the ends of the wire as the source of the studied zero-bias state. Indeed, transport flips the parity of localized modes and suppresses interference, while transport through MZMs preserves island parity and coherent transport~\cite{Sau2015a,Hell2018b}. These localized modes could have been a possible interpretation of the previously observed ZBPs in single-end tunneling experiments~\cite{Moore2018d,prada2019andreev}.

These results suggest that InAs-Al 2DEGs are a promising route towards more complex experiments related to the braiding or fusion of MZMs. We have established coherent transport and parity readout from the transmission phase shifts in Majorana islands, two key results for future topological qubit networks~\cite{Karzig2017b,Plugge2017,Vijay2016}. Future devices will take advantage of improved material quality to allow for increased wire lengths to suppress coherent trivial quasiparticle transport, allowing MZM contributions to be better separated from other potential contributions.

\section*{Methods}
\subsection{Wafer structure.}
The devices were fabricated on wafers grown by molecular beam epitaxy on a InP substrate. The wafer stack consists of a 1~$\rm{\mu m}$ graded $\rm{In_{1-x}Al_{x}As}$ insulating buffer, a $4~\rm{nm}$ $\rm{In_{0.81}Ga_{0.19}As}$ bottom barrier, a $5~\rm{nm}$ InAs quantum well, and a top barrier consisting of $5~\rm{nm}$ $\rm{In_{0.9}Al_{0.1}As}$ for device 1 and $10~\rm{nm}$ $\rm{In_{0.81}Ga_{0.19}As}$ for device 2. A $7~\rm{nm}$ film of epitaxial Al was then grown in-situ without breaking the vacuum of the chamber. The InAs 2DEGs were characterized with a Hall bar geometry (Al removed), which showed a peak mobility of $\mu=17,000~\rm{cm^2V^{-1}s^{-1}}$ for an electron density of $n=1.7\times10^{12}~\rm{cm^{-2}}$ and  $n=7.5\times10^{11}~\rm{cm^{-2}}$ for device 1 and device 2, respectively.
\subsection{Device fabrication.}
Devices were fabricated using standard electron beam lithography techniques. The devices were electrically isolated by etching mesa structures by first removing the top Al film with Al etchant Transene D, followed by a deep III-V chemical wet etch $\rm{H_2 O:C_6 H_8 O_7:H_3 PO_4:H_2 O_2}$ (220:55:3:3). Next, the Al film on the mesa was selectively etched with Al etchant Transene D to produce the Al strip. A $25~\rm{nm}$ thick layer of insulating $\rm{HfO_2}$ was grown by atomic layer deposition at a temperature of $90^{\circ}\mathrm{C}$ over the entire sample. Top gates of Ti/Au (5/25$~\rm{nm}$) were then evaporated and connected to bonding pads with leads of Ti/Au (5/250$~\rm{nm}$).
\subsection{Measurements.}
Electrical measurements were performed by standard lock-in techniques at 166 Hz by applying the sum of a variable dc bias voltage $\vsd$ and an ac excitation voltage of 3 to 10~$\mu \rm{V}$ applied to one of the top ohmic contacts as shown in Fig.~\ref{fig1}a. The resulting current across the device was recorded by grounding a bottom ohmic via a low-impedance current-to-voltage converter, and the four terminal voltage was measured by an ac voltage amplifier with an input impedance of $500~\rm{M\Omega}$. All measurements were taken in a dilution refrigerator with a base temperature of $20~\rm{mK}$ and an electron temperature of $40~\rm{mK}$ estimated by the temperature dependence saturation of ZBP conductance~\cite{Nichele2017}.
\subsection{Data analysis.}
To highlight the oscillating components of the differential conductance $G(\bperp)$, a smooth background was subtracted with a low-degree polynomial Savitzky-Golay filter resulting in $\Delta G$ ~\cite{Savitzky1964a}. Analysis of the oscillations was performed by first performing a fast Fourier transform $F(f)$ of $\Delta G(\bperp)$ using a Hanning window then calculating the power spectral density $PS(f) = |F(f)|^2$. The oscillation amplitude $\Aavg$ was obtained by averaging integrated power spectra. The integration was limited to a band in frequency between $f_1 = 0.55~\rm{mT}^{-1}$ and $f_2 = 0.75~\rm{mT}^{-1}$, spanning the range of a single flux quantum $\Phi_0 = h/e$ through the area $A$ defined by either the central gate $(A_1 = 2.25~\mu\rm{m}^2 )$ or the exterior gates $(A_2 = 3.1~\mu\rm{m}^2)$, where $f = A/\Phi_0$ (see Figs.~\ref{fig2}i-l).

\bibliographystyle{naturemag}

\noindent \textbf{Acknowledgments.} This work was supported by Microsoft Corporation, the Danish National Research Foundation, and the Villum Foundation. We thank Karsten Flensberg, Joshua Folk, Michael Hell, Andrew Higginbotham, Torsten Karzig, Panagiotis Kotetes, Martin Leijnse, Tommy Li, Alice Mahoney, and Ady Stern for useful discussions.

\clearpage

\setcounter{figure}{0}
\renewcommand{\thefigure}{S.\arabic{figure}}
\renewcommand{\theHfigure}{Supplement.\thefigure}

\onecolumngrid
\clearpage
\onecolumngrid

\section{Supplementary Information}

\begin{figure*}[hb]
	\includegraphics[width=0.85\textwidth]{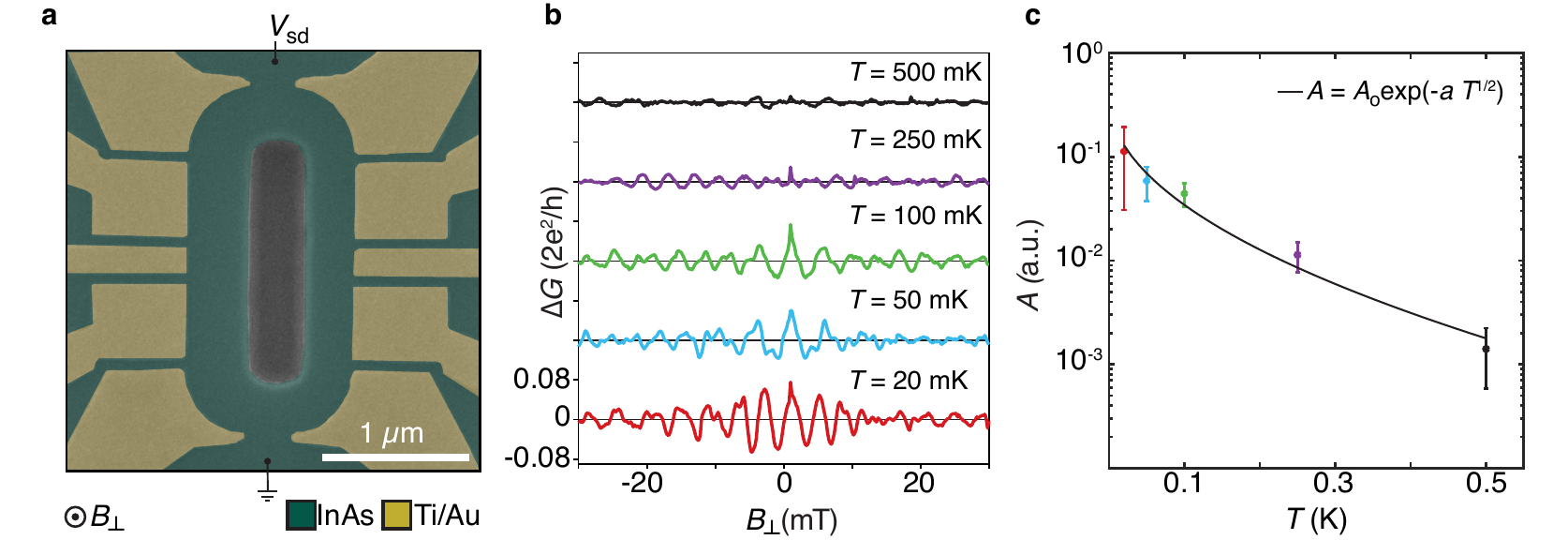}
	\caption{\textbf{Temperature dependence of Aharonov-Bohm oscillations in a normal 2DEG.} \textbf{a}, False-colour electron micrograph of a normal conducting AB interferometer defined in an InAs 2DEG (green) by gates (Ti/Au). The hole forming the interferometer center is created by a wet etch. \textbf{b}, Magnetoconductance $\Delta G$ as a function of perpendicular magnetic field $\bperp$ controlling the flux in the interferometer for several temperatures. Periodic oscillations are observed with a frequency of $f~\sim$ 0.26~mT$^{-1}$, which agrees with a single magnetic flux quantum $h$/$e$ piercing the interferometer loop. \textbf{c}, Temperature dependence of the AB oscillations amplitude $A$ measured from the power spectrum of the curves in \textbf{b}. For a diffusive interferometer, the amplitude $A = A_0\exp(-L/l_\phi(T))$ where $l_\phi(T) \propto T^{-1/2}$ is the phase coherence length and $L = 4.5~\mu$m is the circumference of the interferometer~\cite{Ludwig2004}. The exponential fit $A = A_0\exp(-aT^{1/2})$ gives a base temperature coherence length of $l_\phi(\rm{20~mK}) =$ 4~$\mu$m $\pm~1~\mu$m. Error bars show the standard deviation between 4 data sets at each temperature.
	}
	\label{SI1}
\end{figure*}

\begin{figure}[h]
	\includegraphics[width=0.8\columnwidth]{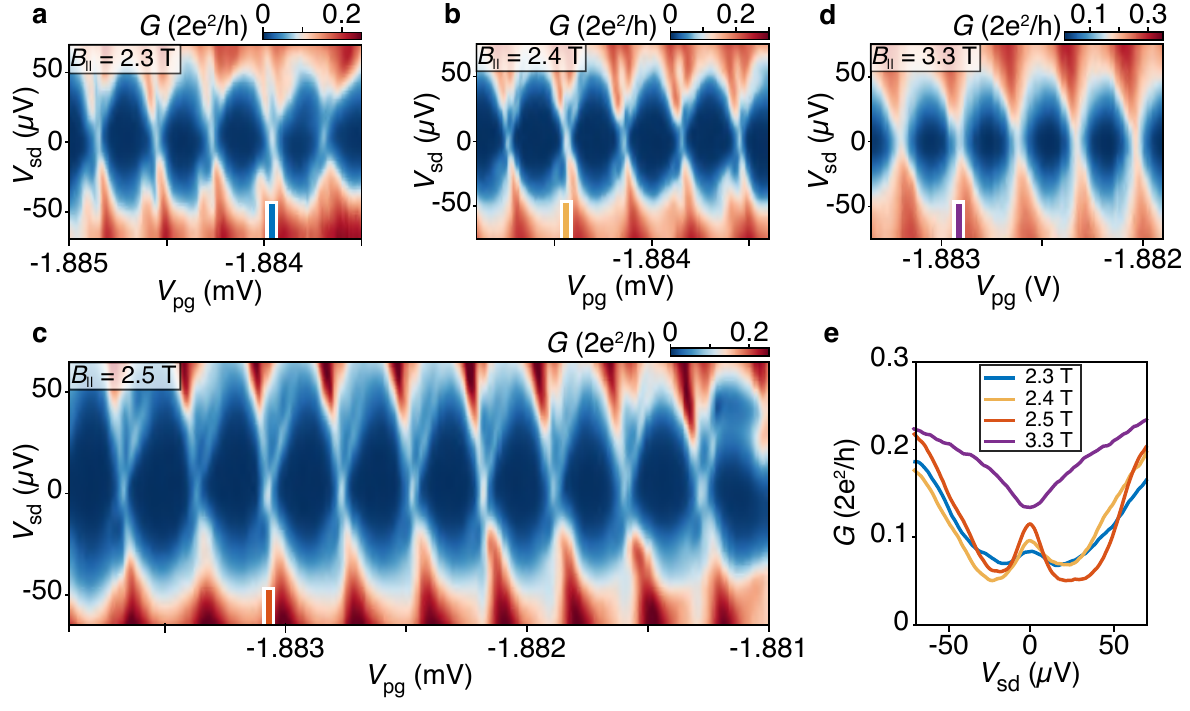}
	\caption{\textbf{Stability of the zero-bias state} \textbf{c-f}, Differential conductance $G$ as a function of $\vsd$ and $\vpg$ showing Coulomb diamonds for $\bpl$ = 2.3 T (\textbf{a}), 2.4 T (\textbf{b}), 2.5 T (\textbf{c}), and 3.3 T (\textbf{d}). \textbf{e}, Line cuts of $G$ vs $\vsd$ at charge degeneracy showing a discrete zero bias peak in the 1$e$ regime and a zero-bias dip in the normal state. The measurements were taken in the same gate configuration as Fig.~1 in the main text}
	\label{SI2}
\end{figure}

\begin{figure*}[h]
	\includegraphics[width=0.75\textwidth]{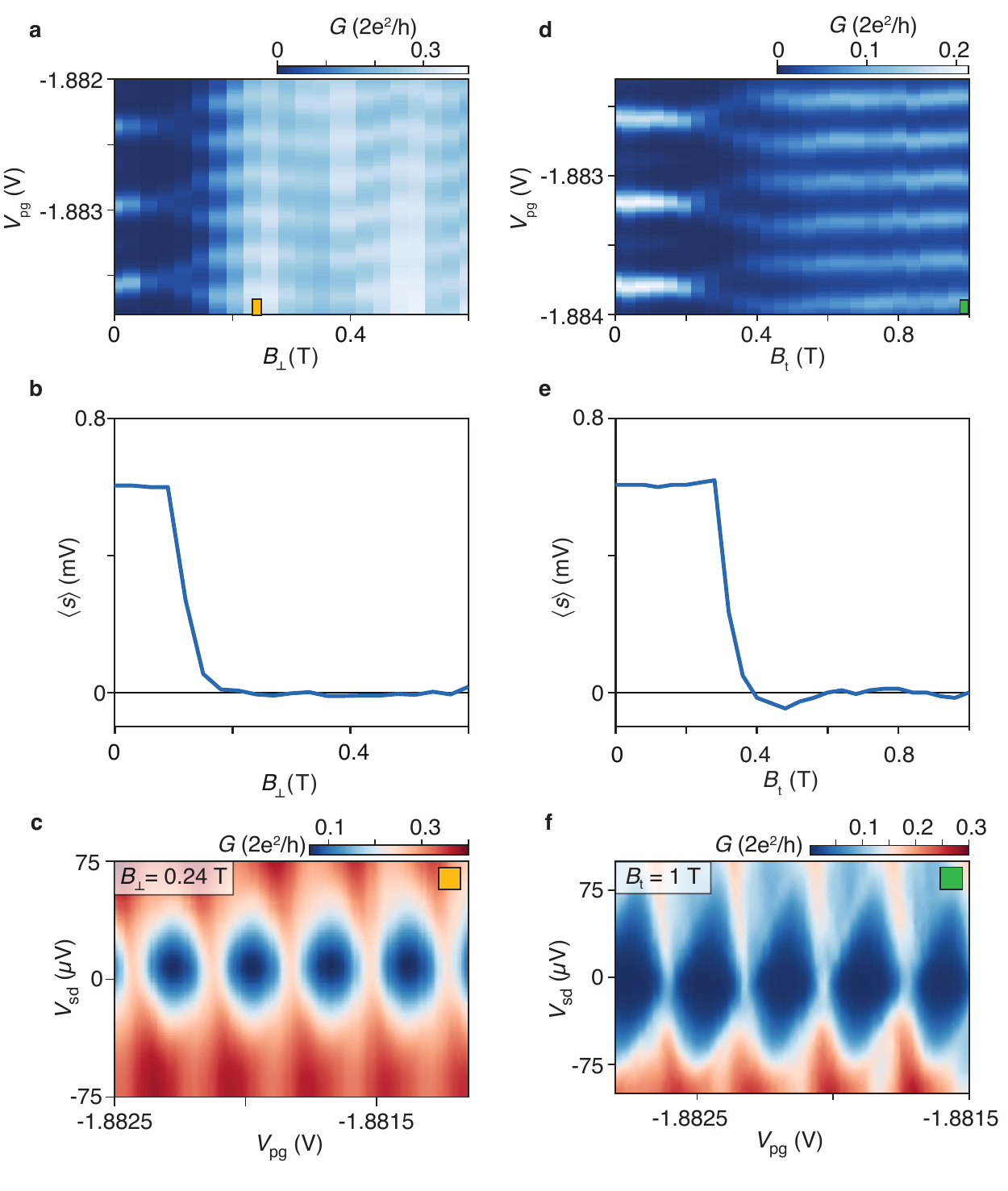}
	\caption{\textbf{Transverse and perpendicular fields for device 1 in the regime of Fig.~1 of the main text}. \textbf{a}, Zero-bias differential conductance $G$ as a function of gate voltage $\vpg$ controlling electron occupancy and perpendicular field, $\bperp$. \textbf{b}, CB peak spacing difference $\Savg$ as a function of $\bperp$. \textbf{c}, Differential conductance $G$ as a function of source-drain bias voltage $\vsd$ and $\vpg$ showing Coulomb diamonds for $\bperp$ = 0.24~T. \textbf{d}, Zero-bias differential conductance $G$ as a function of $\vpg$ and transverse field, $\btr$. \textbf{e}, CB peak spacing difference $\Savg$ as a function of $\btr$. \textbf{f}, Differential conductance $G$ as a function of $\vsd$ and $\vpg$ showing Coulomb diamonds for $\btr$ = 1~T. The field directions are represented in Fig.~1a in the main text.
	}
	\label{SI3}
\end{figure*}

\begin{figure*}
	\includegraphics[width=0.65\textwidth]{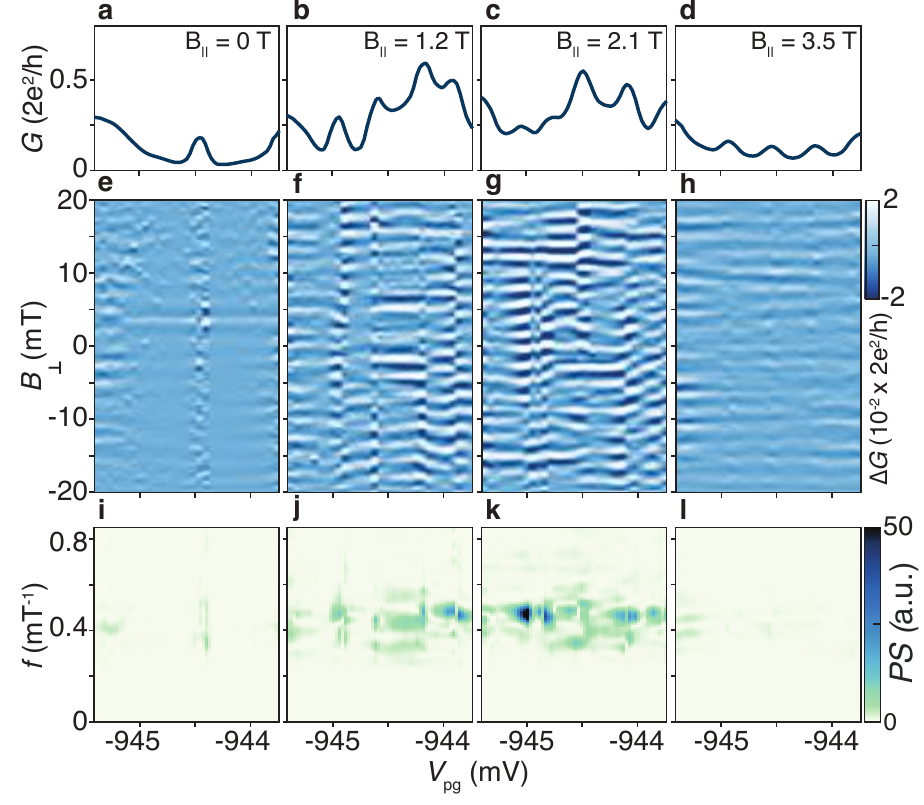}
	\caption{\textbf{Conductance oscillations evolutions in parallel field for Device 2}. Magnetoconductance  for parallel field values $\bpl$ = 0, 1.2, 2.1, and 3.5~T (left to right). \textbf{a-d}, Zero-bias differential conductance $G(\bperp=0$) versus gate voltage $\vpg$ controlling electron occupation. \textbf{e-h}, Magnetoconductance $\Delta G$ as a function of $\vpg$ and perpendicular field $\bperp$ controlling the flux in the interferometer with corresponding power spectra in \textbf{i-l}. A single flux quantum piercing the loop area $A_{\rm loop} \sim 1.8~\mu \rm{m}^2$ corresponds to a frequency $f_{\rm loop} = A_{\rm loop}/(h/e) \sim 0.44$~mT$^{-1}$. \textbf{e-l}, a slowly varying background has been subtracted.
	}
	\label{SI4}
\end{figure*}

\begin{figure*}
	\includegraphics[width=1\textwidth]{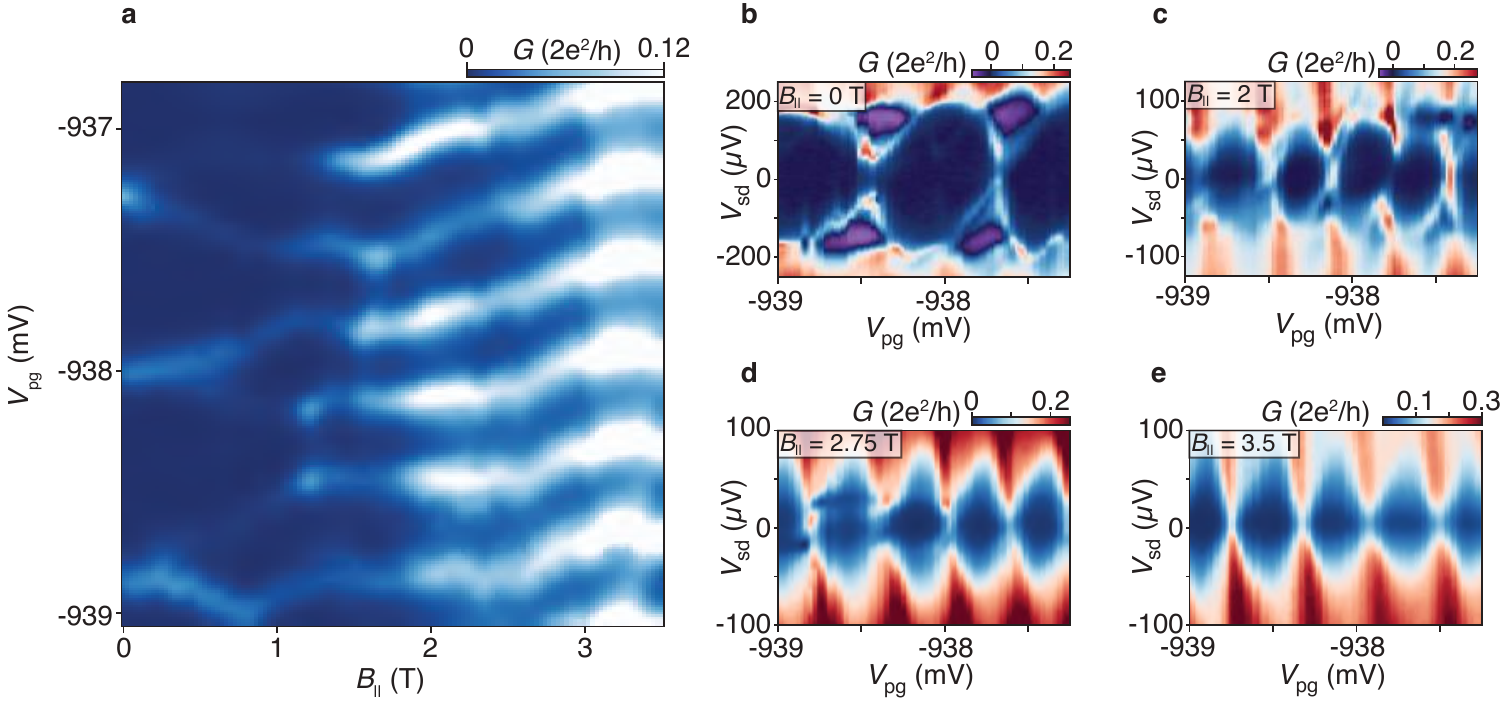}
	\caption{\textbf{Coulomb blockade for device 2.} \textbf{a}, Zero-bias differential conductance $G$ as a function of parallel magnetic field $\bpl$ and gate voltage $\vpg$ controlling the electron occupancy with the reference arm closed. \textbf{b-e}, Differential conductance $G$ as a function of voltage bias $\vsd$ and $\vpg$ for $\bpl$ = 0~T (\textbf{b}), 2~T (\textbf{c}), 2.75~T (\textbf{d}), and 3.5~T (\textbf{e}).
	}
	\label{SI5}
\end{figure*}

\begin{figure*}
	\includegraphics[width=0.9\textwidth]{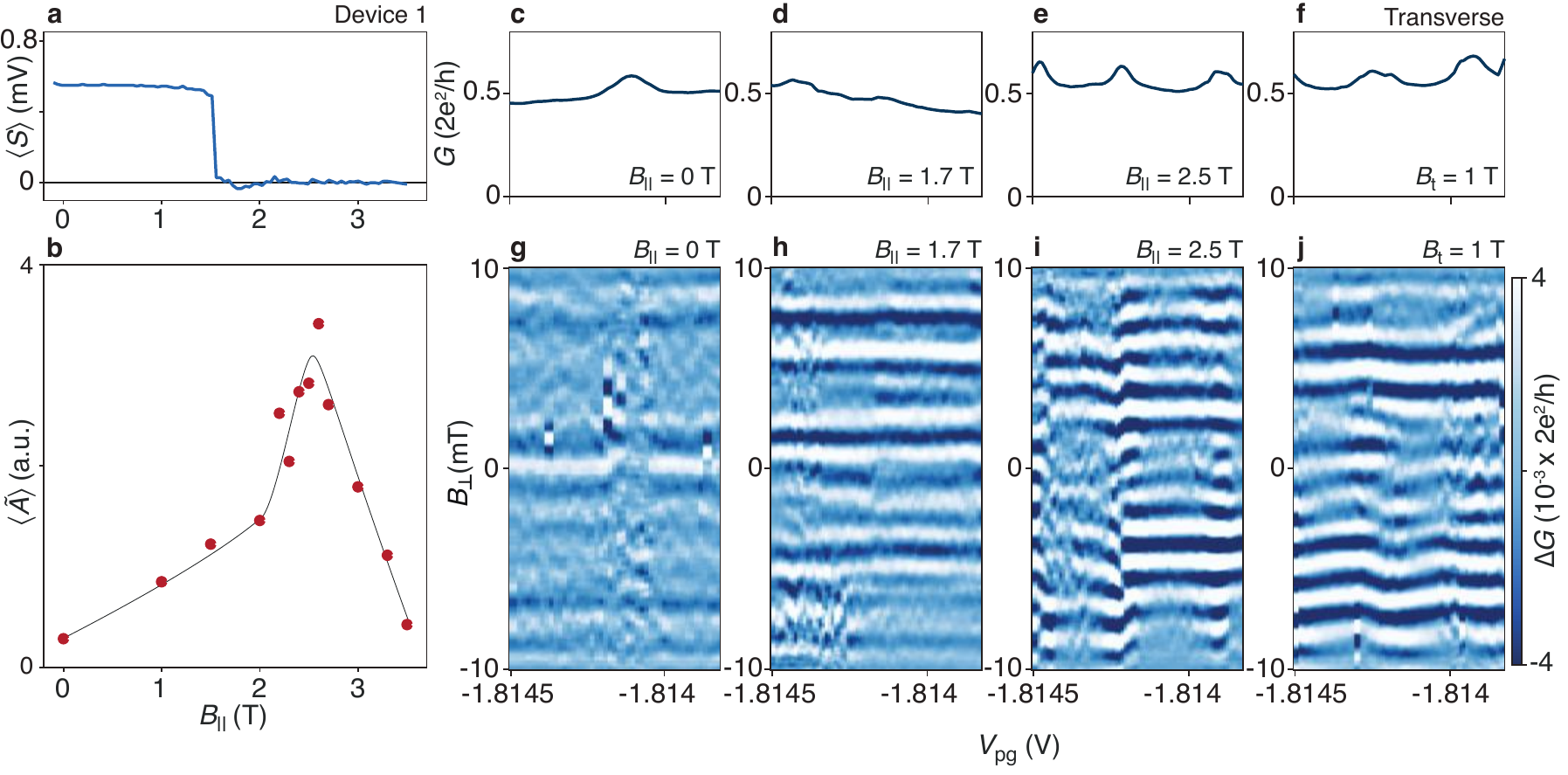}
	\caption{\textbf{Second gate configuration of device 1}. \textbf{a}, Peak spacing difference $\Savg$ as a function of parallel magnetic field $\bpl$. \textbf{b}, Aharonov-Bohm oscillation amplitude $\Aavg$ as a function of $\bpl$. The solid line is a  guide to the eye. \textbf{c-j}, Magnetoconductance for parallel magnetic fields $\bpl$ = 0~T, 1.7~T, and 2.5~T and transverse magnetic field $\btr$ = 1~T (left to right). \textbf{c-f}, Zero-bias differential conductance $G(\bperp=0$) versus gate voltage $\vpg$ used to control electron occupation. \textbf{g-j}, Magnetoconductance $\Delta G$ as a function of $\vpg$ and perpendicular magnetic field $\bperp$ controlling the flux in the interferometer.}
	\label{SI6}
\end{figure*}

\begin{figure}[h]
	\includegraphics[width=0.5\columnwidth]{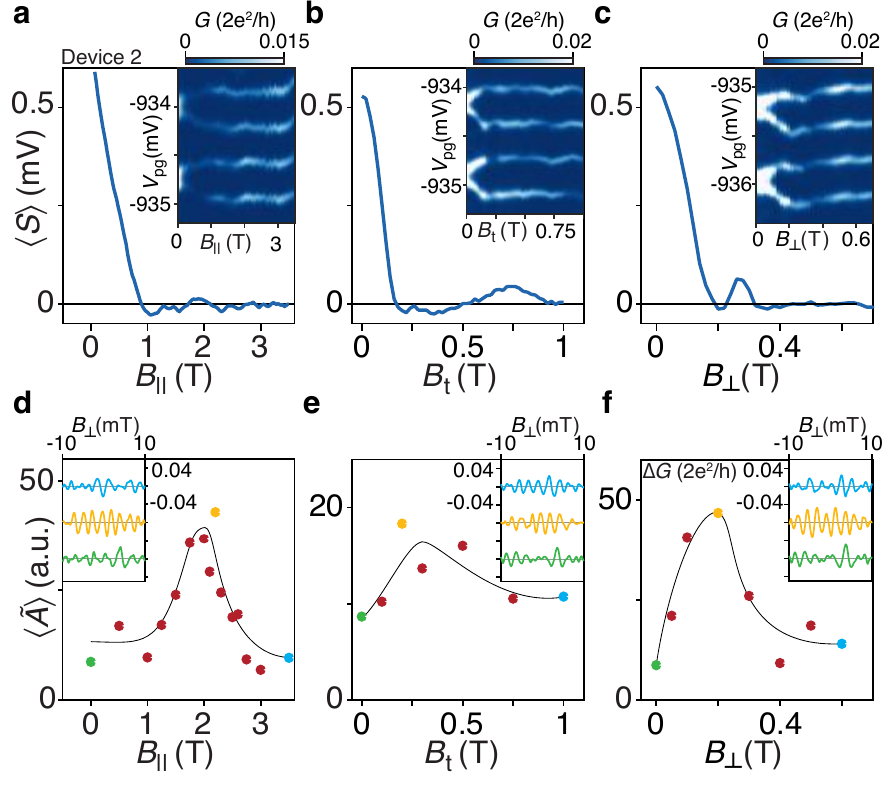}
	\caption{\textbf{Orthogonal magnetic fields.} \textbf{a-c}, Peak spacing difference $\Savg$ as a function of magnetic fields $\bpl$, $\btr$, and $\bperp$ for \textbf{a}, \textbf{b}, and \textbf{c}, respectively. Insets show the zero-bias differential conductance $G$ as a function of magnetic field and gate voltage $\vpg$ controlling electron occupancy with the reference arm closed. \textbf{d-f}, Oscillation amplitude $\Aavg$ as a function of magnetic fields $\bpl$, $\btr$, and $\bperp$ for \textbf{d}, \textbf{e}, and \textbf{f}, respectively. The solid lines are a guide to the eye. Insets show magnetoconductance $\Delta G$ traces as a function of small perpendicular magnetic field $\bperp$. Curves with the largest AB oscillation amplitude are shown for specific magnetic fields indicated by the marker color in the main panel. 
	}
	\label{SI7}
\end{figure}

\end{document}